\documentclass[a4paper,11pt]{article}
\usepackage{jinstpub} 
\usepackage{lineno}
\usepackage{xspace}

\newcommand{\fig}[1]{figure~\ref{fig:#1}}
\newcommand{\sect}[1]{section~\ref{sec:#1}}

\newcommand{\ivalue}{\mbox{I-value}\xspace}

\title{Evaluation of the mean excitation energies of gaseous and liquid argon}

\author{M. Strait}
\affiliation{School of Physics and Astronomy, University of 
Minnesota Twin Cities, Minneapolis, Minnesota 55455, USA}

\emailAdd{straitm@umn.edu}

\newcommand{\garanswerdirectonly}{$(189\pm 8)$\,eV\xspace}

\newcommand{\garanswerwithoutosd}{$(187\pm 6)$\,eV\xspace}

\newcommand{\garanswerosdonly}{$(187\pm 5)$\,eV\xspace}

\newcommand{\garanswer}{$(187\pm 4)$\,eV\xspace}

\newcommand{\laranswer}{$(197\pm 7)$\,eV\xspace}

\abstract{Current and future experiments need to know the stopping power of
liquid argon. It is used directly in calibration, where commonly the
minimum-ionizing portion of muon tracks is used as a standard candle.
Similarly, muon range is used as a measure of muon energy. More broadly, the
stopping power figures into the simulation of all charged particles, and so
uncertainty propagates widely throughout data analysis of all sorts. The main
parameter that controls stopping power is the mean excitation energy, or
I-value. Direct experimental information for argon's I-value come primarily from
measurements of gaseous argon, with a very limited amount of information from
solid argon, and none from liquid argon.  A powerful source of indirect information is also available from
oscillator strength distribution calculations.  We perform a new calculation
and find that from oscillator strength information alone, the I-value of gaseous
argon is $(187\pm 5)$\,eV.  In combination with the direct measurements and other 
calculations, we recommend $(187\pm 4)$\,eV for gaseous argon.
For liquid argon, we evaluate the difference in central value and uncertainty
incurred by the difference of phase and recommend $(197\pm 7)$\,eV.
All uncertainties are given to 68\% C.L. }

\keywords{Interaction of radiation with matter, Neutrino detectors}

\arxivnumber{2212.06286}

\begin{document}
\maketitle
\flushbottom

\frenchspacing

\hyphenation{NBSIR}

\newcommand{\tableerr}
{
  \begin{table}

  \caption{Contributions to the uncertainty on the I-value in the
  OSD calculation, by energy range.  Groups of ranges treated
  as fully correlated are shown without separating horizontal lines
  and with subtotals.  The total 
  uncertainty is found by adding each group in quadrature.}
  \label{tab:osderr}

  \centering

  \begin{tabular}{l c c}

  \hline
  \hline

  Energy & Uncertainty (eV) \\

  \hline

  4s, 4s$'$ (Gibson and Risley) & 0.3 \\

  \hline
  Other discrete below the IP (Chan) & 1.5 \\
  \hline

  Unresolved discrete below the IP (Berkowitz) & 0.2 \\
  \hline

  \hline

  Window resonances 26--29\,eV (Madden, Berrah) & 0.1 \\

  \hline

  1s$\rightarrow$4pm, 3202.3\,eV (Deslattes) & \phantom{0}0.01 \\

  \hline

  15.7596--15.9371 (Berkowitz) & 0.2 \\

  \hline

  15.9371--29.3295 (Samson, Carlson) & 2.7 \\

  29.3295--48.0 (Samson, Carlson) & 0.8 \\

  Subtotal       & 3.6 \\
  \hline

  48.0--79.3 (Watson, Samson, Suzuki) & 0.1 \\

  79.3--243 (Watson, Samson, Suzuki, Henke) & 0.1 \\

  Subtotal & 0.2 \\

  \hline

  243--250 (Suzuki) & \phantom{0}0.02 \\

  243--336 (Suzuki, Chan) & 0.6 \\

  336--500 (Suzuki) & 1.1 \\

  500--929 (Suzuki) & 0.9 \\

  Subtotal & 2.5 \\

  \hline

  929--3202 (Wuilleumier, Zheng, Henke, Suzuki) & 0.8 \\

  3206--10k (Wuilleumier, Zheng, Millar, McCrary) & 1.4 \\

  Subtotal & 2.2 \\

  \hline

  10k--100k (Chantler) & 0.3 \\

  \hline
  \hline

  \end{tabular}

  \end{table}
}

\newcommand{\tablediscrete}
{
  \begin{table}
  \caption{Oscillator strengths for discrete transitions.  The rightmost
  column indicates the primary source of information. Uncertainties are
  judgments made here based on consistency between several measurements 
  or as stated by the original experimental group~\cite{Gibson}\cite{chan1992b}.}
  \label{tab:discrete}
  \centering
  \begin{tabular}{l c c c}

  \hline
  \hline
  Level & Energy (eV) & $f$ & Group \\
  \hline

  4s & 11.62 & $0.0580\pm 0.0034$ & Gibson and Risley\\

  4s$'$ & 11.83 & $0.2214\pm 0.0068$ & Gibson and Risley \\

  5s  & 14.09 & $0.026\pm 0.003$ & Chan \\

  3$\mathrm{\bar d}$ & 14.15 & $0.090\pm 0.009$ & Chan \\

  5s$'$  & 14.26 & $0.012\pm 0.001$ & Chan \\

  3d$'$  & 14.30 & $0.106\pm 0.011$ & Chan \\

  3d &  13.86 & $0.00110\pm 0.00011$ & Chan \\

  4d & 14.71 & $0.0019\pm 0.0002$ & Chan \\

  6s &  14.85 & $0.0144\pm 0.0014$ & Chan \\

  4$\mathrm{\bar d}$ & 14.86 & $0.0484\pm 0.0048$ & Chan \\

  4d$'$ & 15.00 & $0.0209\pm 0.0021$ & Chan \\

  6s$'$ & 15.02 & $0.0221\pm 0.0022$ & Chan \\

  5d & 15.12 & $0.0041\pm 0.0008$ & Chan \\

  7s & 15.19 & $0.0139\pm 0.0014$ & Chan \\

  5$\mathrm{\bar d}$ & 15.19 & $0.043\pm 0.010$ & Chan \\

  Others & 15.5 & $0.18\pm 0.02$ & Berkowitz \\

  \hline
  \hline
  \end{tabular}
  \end{table}
}

\newcommand{\tablecoeff}
{
  \begin{table}

  \caption{Coefficients for the piecewise polynomial fit to various energy
  ranges, used in the OSD calculation.}
  \label{tab:coeff}

  \centering

  \begin{tabular}{l c c c c c c}

  \hline
  \hline

  Energy (eV) & $a_2$ &$a_3$ &$a_4$ &$a_5$ &$a_6$ &$a_7$ \\

  \hline

  15.9371--29.3295\!\! & $-74.283\,0$ & $\phantom{+}386.182$ & $-494.182$ & $-29.413\,5$ & \!$402.690$\! & \!$-187.177$\! \\

  29.3295--48.0 & $\phantom{+}122.781$ & $-890.881$ & $2\,080.83$ & $-1\,516.569$ & --- & --- \\

  48.0--79.3 &  $\phantom{+}16.428\,0$ &  $ -66.036\,0$ &  $ -10.122\,0$ &  $\phantom{+}203.586$ & --- & --- \\ 

  79.3--243 &  $\phantom{+}10.457\,5$ & $\phantom{+}27.350\,8$ & $-512.203$ & $1\,122.41$ & --- & --- \\ 

  336--500 & $\phantom{+}27.194\,2$ & $7\,158.62$ & $-134\,219$ & $\phantom{+}729\,590$ & --- & --- \\

  500--929 & $-36.618\,3$ & $\phantom{+}13\,081.3$ & $-296\,263$ & $1\,967\,090$ & --- & --- \\

  929--3202 & $\phantom{+}31.255\,2$ & $\phantom{+}280.033$ & $\phantom{+}732\,348$ & $-28\,474\,100$ & --- & --- \\

  3206--10k & $-43.245\,8$ & $\phantom{+}207\,562$ & \!$-39\,331\,400$\! & \!\!$3\,787\,040\,000$\!\! & --- & --- \\

  \hline
  \hline

  \end{tabular}

  \end{table}
}

\newcommand{\figureone}
{
\begin{figure}
  \centering

  \includegraphics[width=0.64\columnwidth]{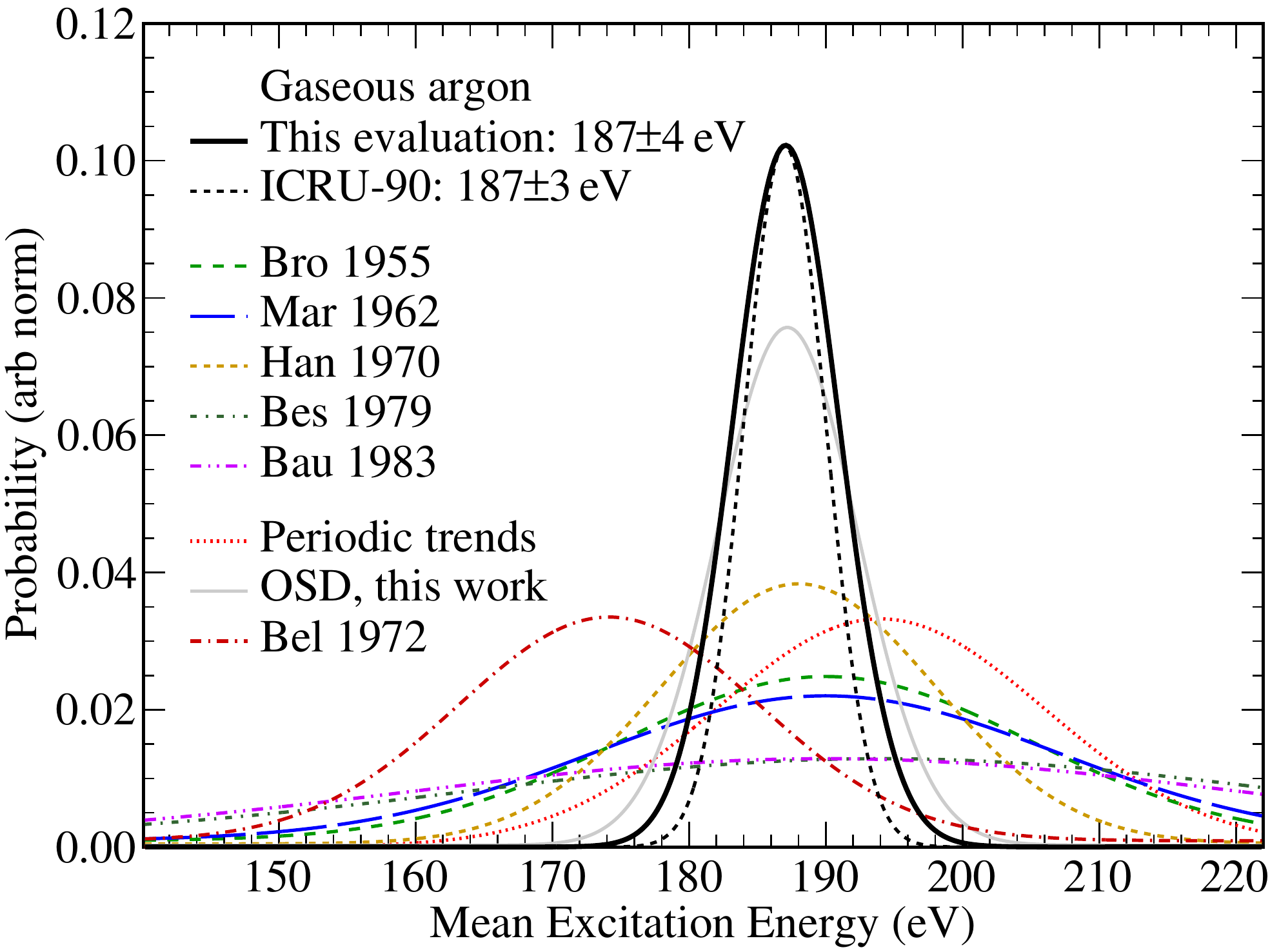}

  \caption{Evaluation of the mean excitation energy of gaseous argon, using
     experimental results~\cite{Brolley,Martin,Hanke,Baumgart,Besenbacher1979},
     interpolation from periodic trends, and 
     calculations~\cite{bell,Kamakura}. The ICRU-90 recommendation is shown for
     comparison, interpreting their central value and uncertainty as a Gaussian PDF.
     All curves except for the one for ICRU-90 share a normalization.}

  \label{fig:myeval}
\end{figure}
}

\newcommand{\figuretwo}
{
\begin{figure}
  \centering

  \includegraphics[width=0.64\columnwidth]{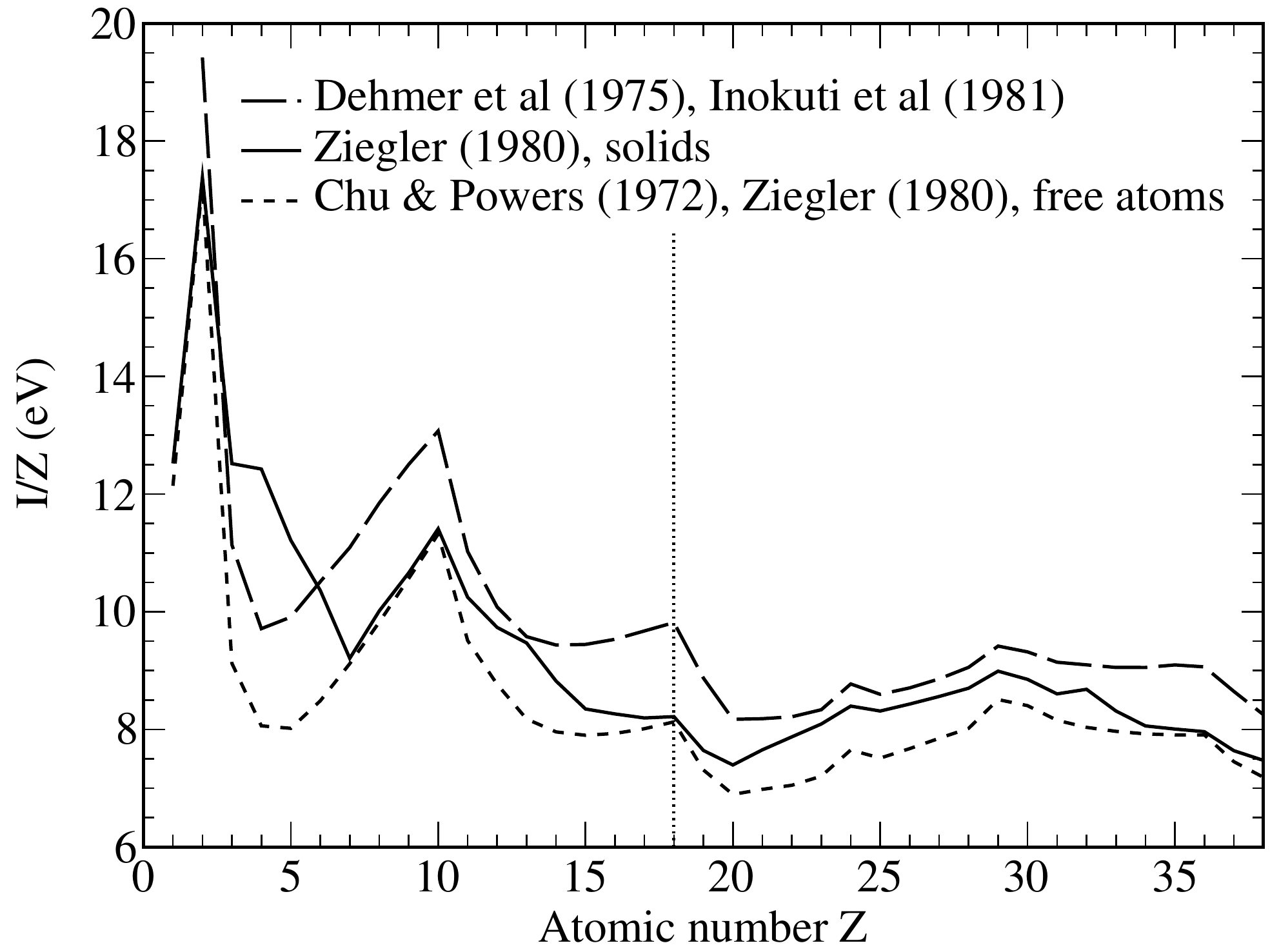}

  \caption{Reproduction of ICRU-37 Fig~3.2 for $Z = 1$--38. Note suppressed zero.  The vertical
    dashed line indicates argon's atomic number.}

  \label{fig:calci}
\end{figure}
}

\newcommand{\osdfigscale}{0.69\columnwidth}

\newcommand{\figurethree}
{
\begin{figure}
  \centering \includegraphics[width=\osdfigscale]{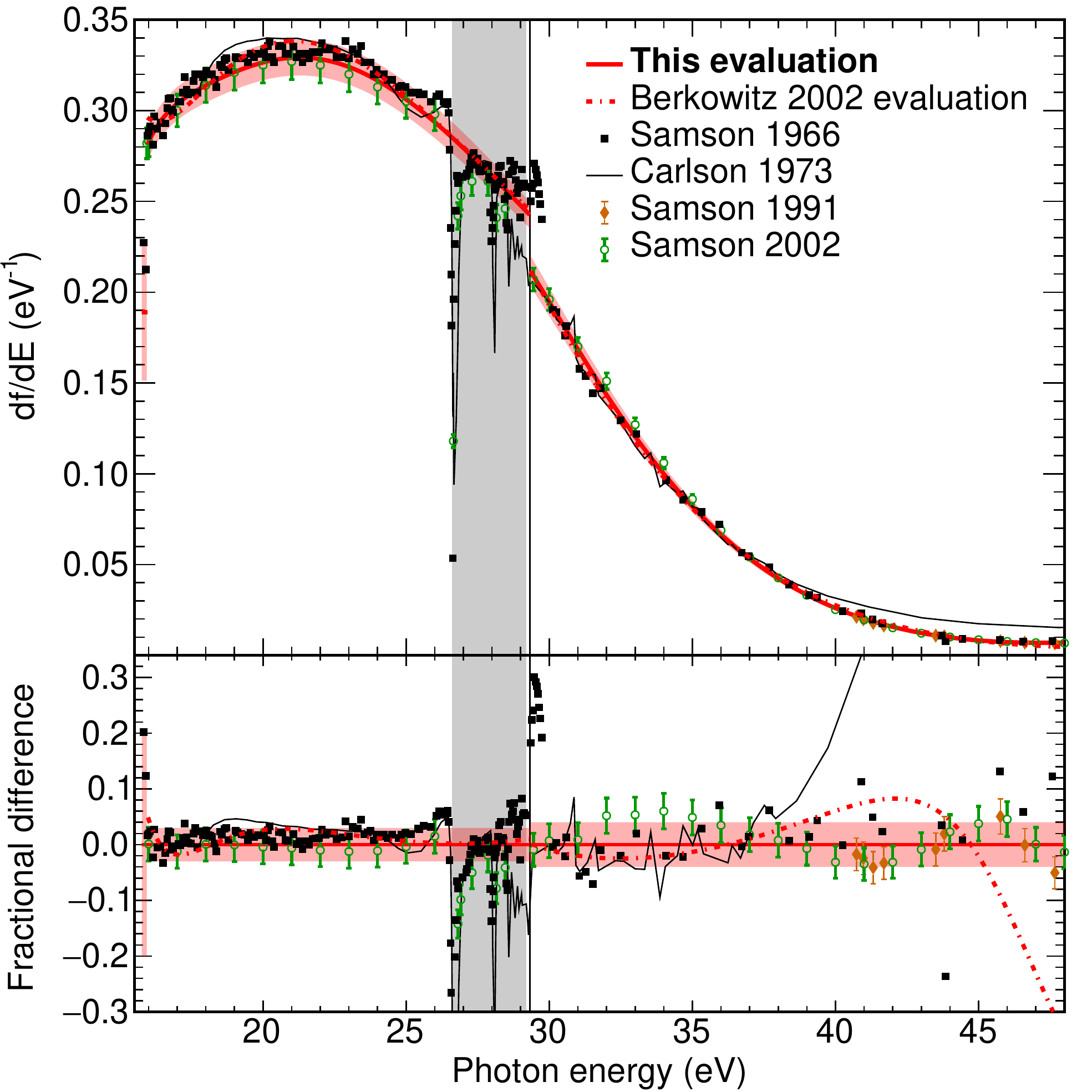}

  \caption{Oscillator strength distribution for gaseous argon, 15--48\,eV,
showing data of Samson 1966~\cite{samson1966}, Carlson et al
1973~\cite{carlson1973}, Samson et al 1991~\cite{samson1991}, and Samson \&
Stolte 2002~\cite{samson2002}. This note's evaluation is shown as the solid red
line and Berkowitz 2002's evaluation~\cite{Berkowitz} in dot-dashed red. The shaded energy range
from 26.6 to 29.2\,eV is handled specially; see the text.  The vertical line
at 29.3295\,eV is the boundary between polynomial fits used in the evaluations.
The bottom pane, on this plot and the following plots, shows the fractional differences between the present evaluation and
the various data and other evaluations.}

  \label{fig:osd3}
\end{figure}
}

\newcommand{\figurefour}
{
\begin{figure}
  \centering \includegraphics[width=\osdfigscale]{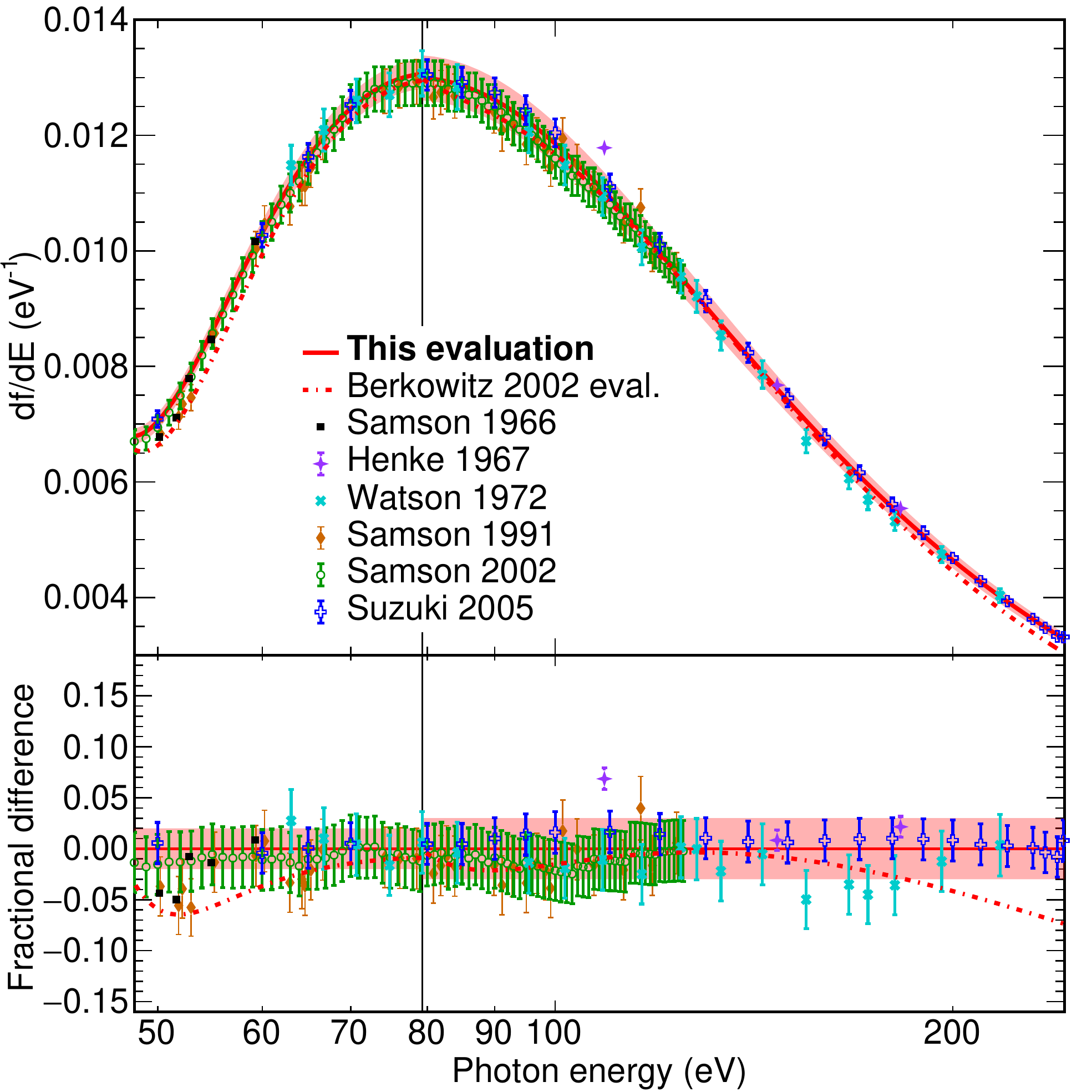}

  \caption{Oscillator strength distribution for gaseous argon, 48--230\,eV,
showing the data of Samson 1966~\cite{samson1966}, Henke et al
1967~\cite{henke1967}, Watson 1972~\cite{Watson}, Samson et al
1991~\cite{samson1991}, Samson \& Stolte 2002~\cite{samson2002}, and Suzuki \& Saito 2005~\cite{suzuki2005}.  The evaluation of Berkowitz 2002~\cite{Berkowitz} is also shown. The conventions are the same as for \fig{osd3}. Note suppressed
zero.}

  \label{fig:osd4}
\end{figure}
}

\newcommand{\figurefive}
{
\begin{figure}
  \centering \includegraphics[width=\osdfigscale]{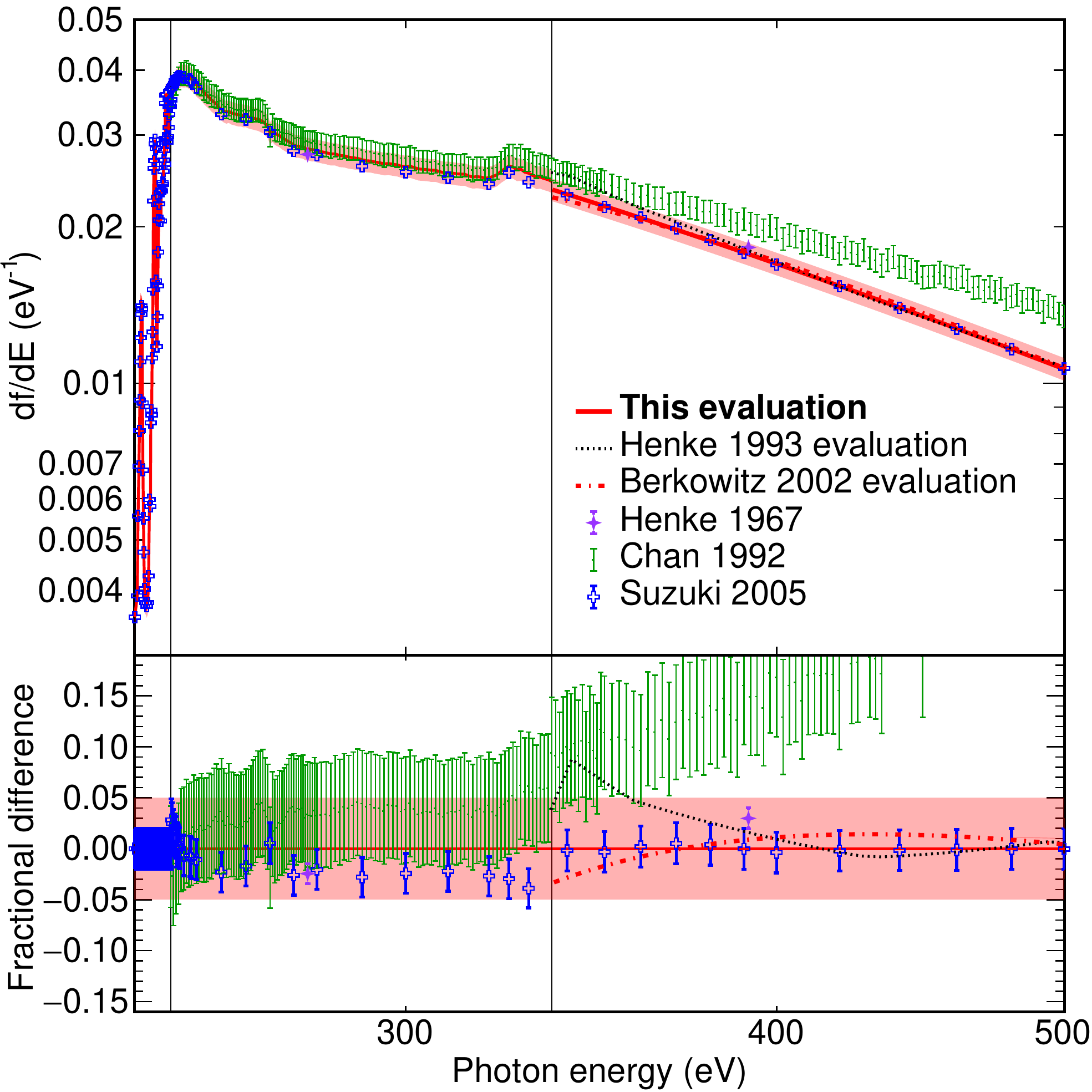}

  \caption{Oscillator strength distribution for gaseous argon, 230--500\,eV,
showing the data of Henke et al 1967~\cite{henke1967}, Chan et al
1992~\cite{chan1992b}, and Suzuki \& Saito 2005~\cite{suzuki2005}, as well as
the evaluations of Henke et al 1993~\cite{henke1993} and Berkowitz 2002~\cite{Berkowitz}. Between 230 and 336\,eV, Berkowitz uses the
data of Chan 1992 without a functional form. }

  \label{fig:osd5}
\end{figure}
}

\newcommand{\figuresix}
{
\begin{figure}
  \centering \includegraphics[width=\osdfigscale]{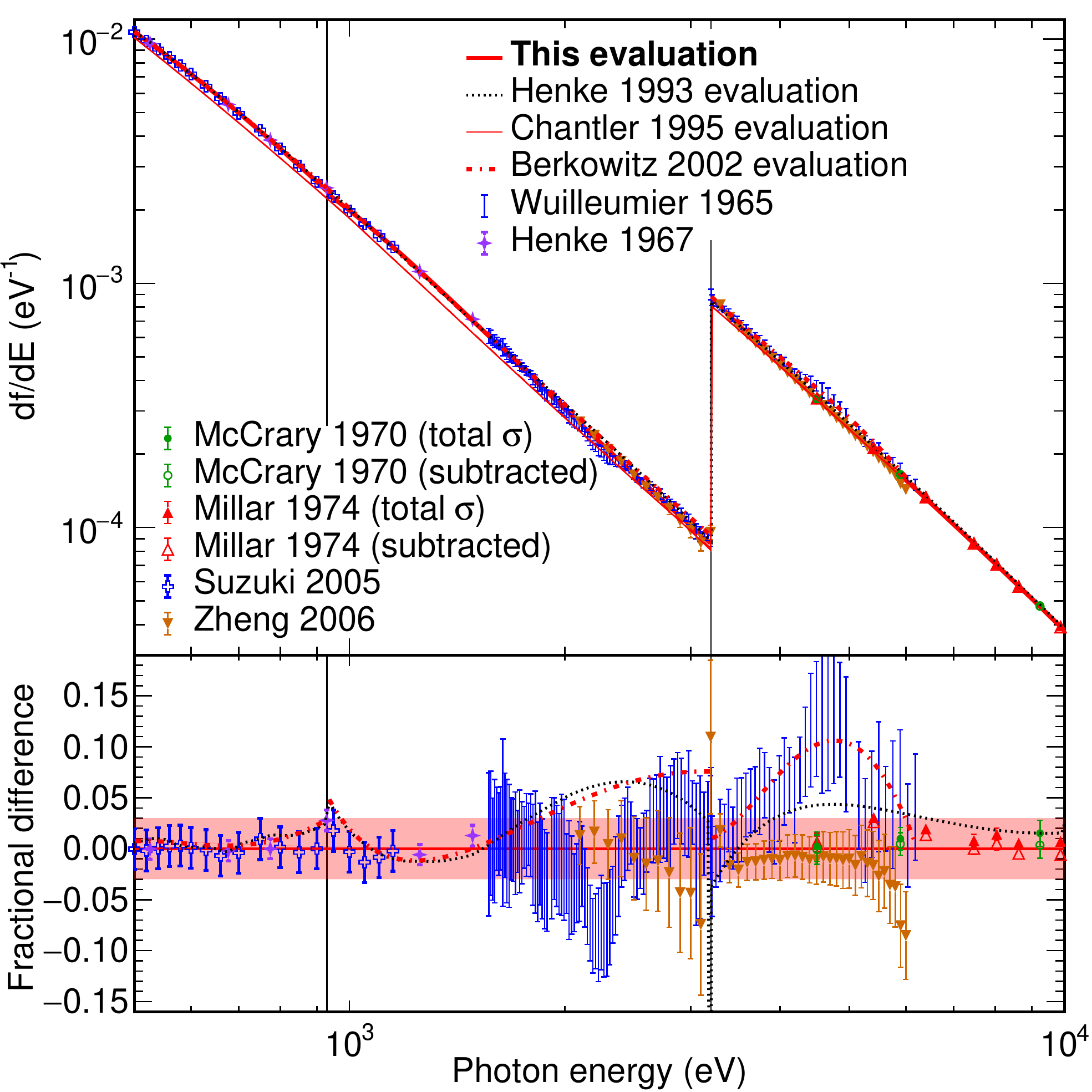}

  \caption{Oscillator strength distribution for gaseous argon, 
500--10,000\,eV, with data of
Wuilleumier 1965~\cite{Wuilleumier},
Henke et al 1967~\cite{henke1967},
McCrary et al 1970~\cite{McCrary},
Millar \& Greening 1974~\cite{millar1974},
Suzuki \& Saito 2005~\cite{suzuki2005},
and
Zheng et al 2006~\cite{zheng2006},
and the evaluations of Henke et al 1993~\cite{henke1993}, Chantler 1995~\cite{Chantler}, and Berkowitz 2002~\cite{Berkowitz}. Chantler 1993 is divided into the total cross section, which is what
   attenuation experiments measure, and the photoabsorption cross section,
   which is of interest for the OSD calculation. The difference is used to
  correct the attenuation measurements of McCrary and Millar.}

  \label{fig:osd6}
\end{figure}
}

\newcommand{\figureseven}
{
\begin{figure}
  \centering \includegraphics[width=\osdfigscale]{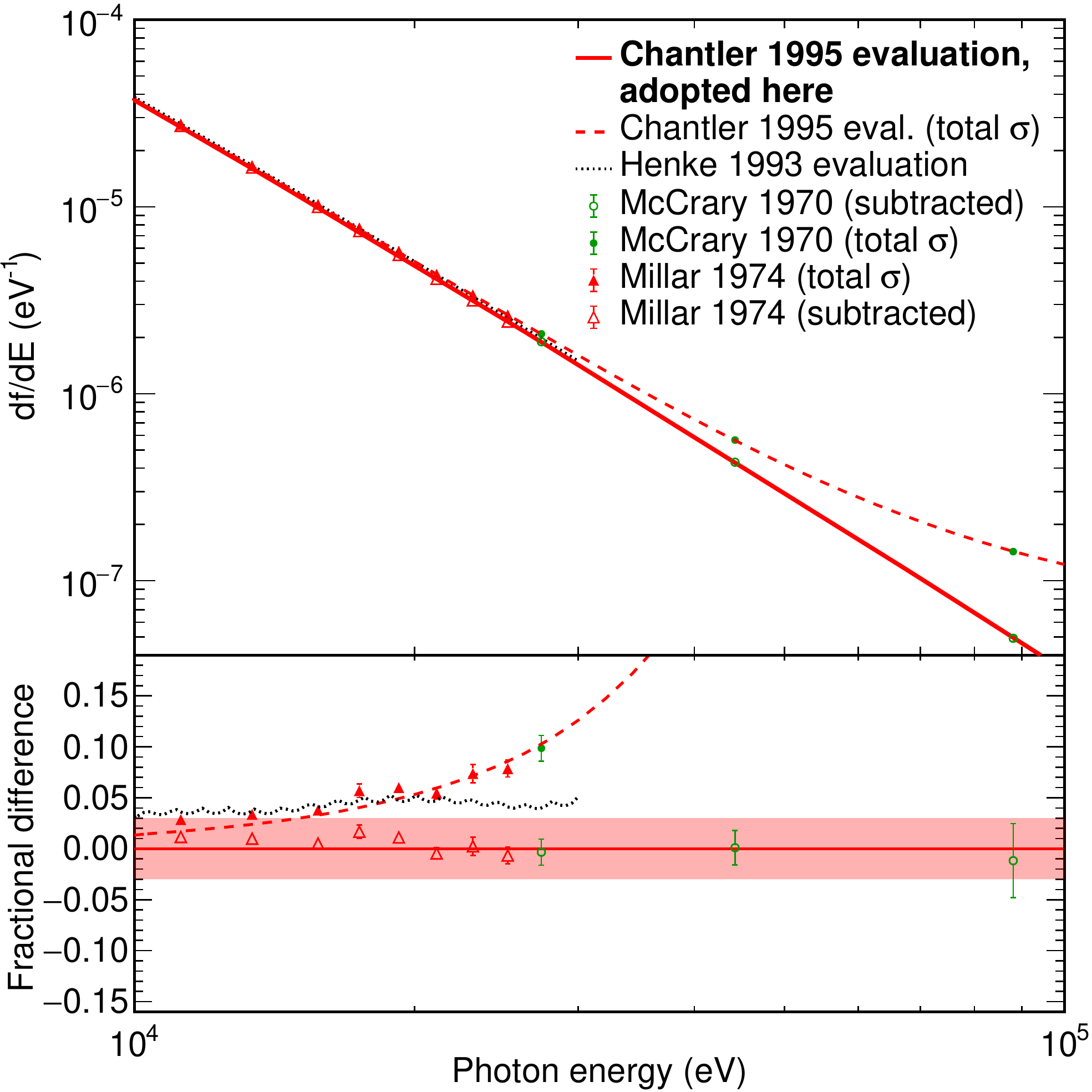}%

  \caption{Oscillator strength distribution for gaseous argon from $10^4$ to
   $10^5$\,eV.  The data of McCrary et al 1970~\cite{McCrary}, and
   Millar \& Greening 1974~\cite{millar1974} are shown, along with
the evaluations of Henke 1993~\cite{henke1993} and Chantler~\cite{Chantler}.
   Subtraction to obtain oscillator strength from attenuation measurements is done as in \fig{osd6}.  The Chantler evaluation is
   used directly in the present evaluation.
   }

  \label{fig:osd7}
\end{figure}
}

\newcommand{\figurephase}
{
\begin{figure}
  \centering

  \includegraphics[width=0.64\columnwidth]{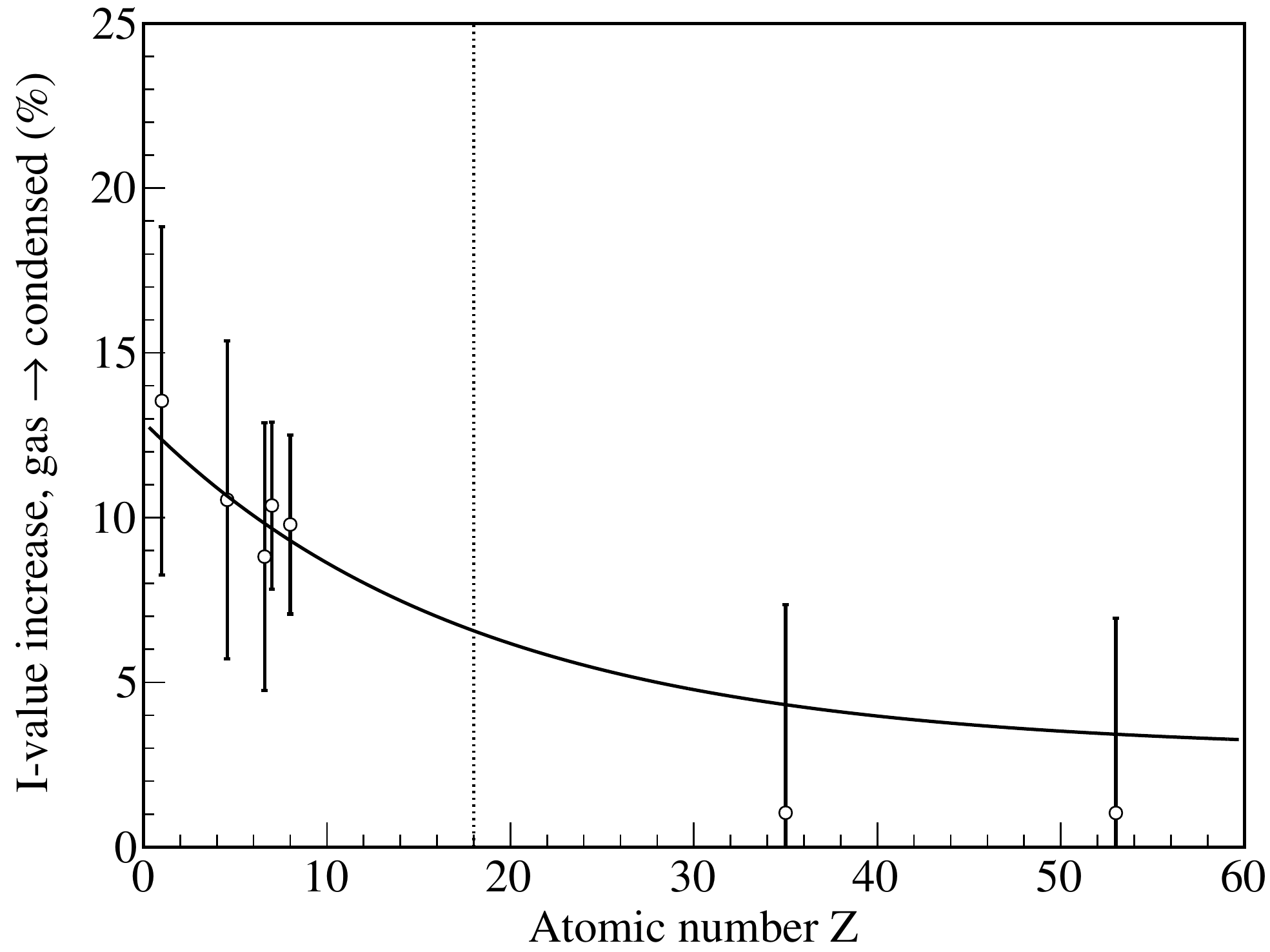}

  \caption{Effect of phase on I-value.  The points with error bars are each discussed
       in the text.  For compounds water and the hydrocarbons, $Z$ is a weighted average of the
       constituent elements.  The atomic number of argon is shown as a vertical dotted line.
       A smooth curve is drawn to give an estimate of the effect for argon.}

  \label{fig:interp}
\end{figure}
}

\title{Evaluation of the mean excitation energy of liquid argon}

\author{M.~Strait}
\thanks{straitm@umn.edu}
\affiliation{School of Physics and Astronomy, University of 
Minnesota Twin Cities, Minneapolis, Minnesota 55455, USA}


\section{Introduction}

The International Commission on Radiation Units and Measurements (ICRU) recommended
values and uncertainties for the mean excitation energy of gaseous argon twice, once in 1984~\cite{icru37}
and again in 2016~\cite{icru90}.  The first of these was in Report 37 which gives an evaluated value of $(188\pm 10$)\,eV.
The uncertainty is not at 68\% C.L., but is a ``[figure] of merit,
arrived at by subjective judgments, and with a meaning that is not easily
defined."  The report further explains that errors are given at roughly 90\% C.L., and that
one ``could convert them to `standard deviations' by multiplying them by a
factor of about one half.''  These two statements indicate difference confidence levels. 
However, ICRU-37 derives from NBSIR 82-2550~\cite{nbsir}, which
gives only the first of these uncertainty interpretations, so this note will consider
ICRU-37 uncertainties to be at 90\% C.L.; this also agrees with ICRU-90's comments on ICRU-37.
The 68\% C.L. uncertainty is therefore 6\,eV.

ICRU-37 uses four experimental results for stopping power and
range~\cite{Brolley,Martin,Hanke,Besenbacher1979} in their evaluation.  
Other methods of estimating I-values are cited, such as the
semi-empirical oscillator strength distribution calculated in
Ref.~\cite{Eggarter}, but they are not used in the evaluation.

ICRU Report 90~\cite{icru90} updates the I-value of
gaseous argon to $(187\pm 3)$\,eV, where the uncertainty now means roughly the
usual one standard deviation (``an interval having a confidence of
approximately 67\%'').  The only new experimental result used is
Ref.~\cite{Baumgart}, which does not add much.  The decrease in the
recommended uncertainty is almost entirely from inclusion of oscillator strength
distribution calculations.

This note re-evaluates the sources of information on gaseous argon's I-value.
First, direct experimental evidence from stopping power and range measurements
are reviewed in \sect{expgas}.  Second, the state of oscillator strength
distribution calculations is reviewed, and a new calculation performed, in
\sect{calc}.  Other indirect methods of estimating the I-value are also
reviewed in this section.  In \sect{gaseval} all of the information on the
I-value of gaseous argon is combined into a new recommended value and
uncertainty.  In the following \sect{liquideval}, information on phase
effects are reviewed and we evaluate an I-value for liquid argon.  Finally, the
major implications of the recommended I-value for liquid argon for an
experiment such as DUNE~\cite{DUNEtdr} are given in \sect{implications}.

\section{\label{sec:expgas}Gaseous argon experiments}

\subsection{Brolley \& Ribe 1955}

Brolley \& Ribe~\cite{Brolley} measured the stopping power of deuterons in argon
gas relative to air.  Deuterons with initial energy 10.05\,MeV were sent through
a cell filled with argon, and the pressure in the cell was adjusted until a 
downstream NaI(Tl) crystal registered a drop of 25\% of the initial energy.
This is a stopping power measurement that samples the $dE/dx$ only for fast
projectiles.  It therefore suffers less than a range measurement from
uncertainties related to slow particles, primarily in the difficulty of
evaluating shell corrections.

The authors do not directly report an I-value for argon, but rather quote an
``absolute stopping cross section $dE/dx$'' of
$(3.72\pm 0.08)\times 10^{-15}\,\mathrm{eV\,cm}^2$ for argon gas at 219.2~mmHg,
i.e.  $56.1\,\mathrm{MeV\,g^{-1}\,cm^{2}}$.  From this, the ICRU inferred an
I-value of $(190\pm 15)$\,eV.  The uncertainty in the cross section appears to have been evaluated from a combination of
the
degree of consistency between runs at the same argon pressure, 1.4\%, and the
uncertainty of the initial beam energy, 1.3\%.

The authors are not clear about the confidence level of their uncertainties.
A reanalysis shows that their $\pm 0.08\times 10^{-15}\,\mathrm{eV\,cm}^2$ directly
corresponds to ICRU's $\pm 15$\,eV.  Given that the three uncertainties quoted by the ICRU
for Refs.~\cite{Brolley,Martin,Hanke} are $\pm 15$\,eV, $\pm 7$\,eV and
$\pm 10$\,eV, that these measurements have consistent central values, and that
the final ICRU recommendation is $10$\,eV at 90\% C.L. (``subjective''), it
appears that the ICRU has assumed that Brolley \& Ribe's
$\pm 0.08\times 10^{-15}\,\mathrm{eV\,cm}^2$ is at roughly 68\% C.L.  This is the most
conservative likely interpretation (i.e. the other sensible choices are 90\% or 95\% confidence, and these
would indicate smaller uncertainties), so the same will be assumed in this note's evaluation.

\subsection{Martin \& Northcliffe 1962}

Martin \& Northcliffe~\cite{Martin} measured energy loss of few-MeV alpha
particles in gaseous argon and report an I-value of $(190\pm 17)$\,eV.  The
ICRU-37 table shows this as $(190\pm 7)$\,eV instead, as does
predecessor document NBSIR 82-2550. As Ref.~\cite{Martin}
unambiguously says ``$\pm 17\,$eV'' twice, this appears to be a simple mistake,
not a re-evaluation.  With this misreading, Martin \& Northcliffe would be the
best experimental result, while the correct uncertainty makes it subdominant.

The authors give a detailed list of sources of uncertainty as well as a
discussion of theoretical difficulties in calculating corrections at low energy needed
to obtain the I-value for argon.  The confidence level of their result would seem to be
subjective --- ``obtained by estimating the maximum and minimum slopes
consistent with the errors displayed.''  This note's evaluation treats the uncertainty as
being at 68\% C.L., but this may be very conservative.  Like the previous experiment, this is a
stopping power measurement and is relatively less vulnerable to uncertainties related to very slow particles.

\subsection{Hanke \& Bichsel 1970}

In another stopping power measurement, Hanke \& Bichsel~\cite{Hanke} used alpha particles from radioactive decay to
measure the I-value of gaseous argon.  The authors quote 182\,eV and 167\,eV as
their results, for two evaluations of shell corrections.  ICRU-37 uses a
re-evaluation of $(188\pm 10)$\,eV.  Hanke \& Bichsel provides the most
information on gaseous argon's I-value (given that we have used the correct
uncertainty of 17\,eV from Ref.~\cite{Martin}).  From context, the ICRU quoted uncertainty is
probably meant to be at 68\% C.L., but unfortunately it is not clear how it
was obtained.

\subsection{Besenbacher et al 1979}

Ref.~\cite{Besenbacher1979} reports on a measurement of stopping power for protons
in the range 40\,keV to 1\,MeV and alpha particles in the range 100\,keV to
2.4\,MeV.  The I-value of 194\,eV is quoted with no error in their Table~II.
ICRU has not evaluated an uncertainty either, nor is it clear how it would be
done.  At such low energies, shell corrections and other complications are very
important.  For the purpose of this note's evaluation, it is assumed that
this experiment is a factor of several less precise than those that report
uncertainties, and an error of $\pm 30$\,eV has been assigned.  The final result
below is insensitive to the precise value of this error; so long as it is several
times larger than the uncertainty of the more reliable inputs, this note's
evaluation of the central value and uncertainty are both unchanged to within
the precision displayed.

\subsection{Baumgart et al 1983}

Ref.~\cite{Baumgart} (not used in ICRU-37) reports on a measurement of the
stopping power of argon to protons of between 60 and 800\,keV.  Similarly to
the previous experiment, a value of 190\,eV is given, but no uncertainty is
quoted, and it is not clear how one would be extracted.  Shell corrections are,
again, a major concern.  An error of $\pm 30$\,eV has again been assigned for
this note's evaluation.

\section{\label{sec:calc} Calculations for gaseous argon}

\subsection{Oscillator strength distribution}


ICRU-37 says that the most reliable calculations of I-value come from
semi-empirical dipole oscillator strength distributions (OSD), i.e. the photoabsorption
cross section as a function of energy.  ICRU-37 does not use OSD calculations
as part of its evaluation of the recommended I-value for argon, but ICRU-90
does, with much of the reason for its small recommended uncertainty of 3\,eV
being its adopted uncertainty of 2\,eV for the calculation of Kamakura et al
2006~\cite{Kamakura}, who reported 191\,eV.  However, neither this reference,
nor the older calculations from Kumar \& Meath 1985~\cite{Kumar_Meath} and
Eggarter 1975~\cite{Eggarter}, report an uncertainty themselves.
ICRU-90 explains its own by saying that it is ``based on those quoted for
similar results.''

This is on very shaky ground.  Similar results would only have similar
uncertainties if the uncertainties in the underlying photon cross section data
were similar between the various materials.  But the underlying data for
Kamakura come from the review of Berkowitz 2002~\cite{Berkowitz}, who cautions that
``Information on the oscillation strengths [of argon] is still rather
limited'', a warning that does not appear for similar cases (e.g.  O,
$\mathrm{O_2}$, N, $\mathrm{N}_2$, Ne).  
Moreover, ICRU-90 averages several OSD calculations,
then expands all the errors such that the reduced $\chi^2$ is unity.  But the
calculations are not independent, being based on mostly the same underlying
data, so they cannot be validly combined in this way.

The calculation of Kamakura is based on the recommended
oscillator strengths from Berkowitz.  Here we will
repeat the calculation, adding more recent experimental data, and estimating
uncertainties.  The result is \garanswerosdonly, and is found by \[\log I =
\left.\left(\sum_n f_n \log(E_n) + \int_\mathrm{IP}^\infty \frac{df}{dE}
\log(E) dE\right)\middle/S(0)\right.,\] where \[S(0) = \sum_n f_n +
\int_\mathrm{IP}^\infty \frac{df}{dE} dE,\] and in each case the sum is over
discrete states and the integral is over the continuum from the ionization
potential to infinity.  $E$ is the incoming photon energy and $f$ is the
oscillator strength, i.e. \[f = \frac{2 \epsilon_0 m_e c} {\pi e^2
\hbar}\sigma,\] where $\sigma$ is the photoabsorption cross section,
$\epsilon_0$ is the permittivity of free space, $m_e$ is the mass of the
electron, $c$ is the speed of light, and $e$ is the elementary charge.

By the Thomas-Reiche-Kuhn sum rule, $S(0) = Z$, i.e. 18 for argon, up to
relativistic and multipole effects that are expected to be negligible for
low-$Z$ elements.  We find $S(0)$ to be $17.8\pm0.8$, in good agreement
with the sum rule.  No explicit correction to the oscillator
strengths to make $S(0)$ equal
to 18 is made, as the I-value is invariant under a uniform scaling of
oscillator strengths, and we do not impose additional constraints that would
motivate any non-uniform scaling.

The following discussion examines energy ranges from lowest to highest.
Our method for adopting central values for each part of the spectrum will closely follow
the choices and methods of Berkowitz's review, updated with newer data.

\tablediscrete

\tableerr

First, the discrete spectrum.  The values listed in table~\ref{tab:discrete}
were chosen.  The uncertainties adopted are primarily those stated by the
experimental groups. As an exception, the uncertainties for the 5d and
5$\mathrm{\bar d}$ levels are expanded by a factor of two because Berkowitz
finds the values suspiciously large.  Within each of the two experimental
groups, Chan et al~\cite{chan1992b} and Gibson \& Risley~\cite{Gibson}, we
conservatively take the uncertainties to be fully correlated.  The final
entry in the table is Berkowitz's estimate for all unresolved discrete
transitions close to the ionization potential.  Two uncertainties are
assigned to Berkowitz's estimate, a 10\% uncertainty fully correlated to the discrete
transitions attributed to Chan, since Berkowitz's number is an extrapolation based
primarily on Chan's measurements, and an uncorrelated 10\% uncertainty to 
cover the extrapolation procedure.  Despite the conservatism used in these
decisions,
the overall uncertainty to the I-value from transitions below the ionization
potential, shown in table~\ref{tab:osderr}, is only $\pm 1.5$\,eV, which is a minor contributor to the total.

\figurethree

For the narrow energy range between the $\mathrm{^2 P_{3/2}}$ and $\mathrm{^2
P_{1/2}}$ ionization potentials, Berkowitz assumes a constant cross section
of 20.75\,Mb.  We accept this and assign a 20\% uncertainty to this cross
section to cover the data of Samson 1966~\cite{samson1966} as shown in
\fig{osd3}.  This section does not contribute significantly to the
I-value or its uncertainty.

Following Berkowitz, the continuum is partitioned into several energy
ranges, and the oscillator strength data fit to polynomials within
these ranges.  These are of the form \[\mathrm{\frac{df}{dE} = \frac{eV}{Ry}}
\sum_{i = 2}^7 a_i y^i,\] where $\mathrm{eV/Ry} = 1/13.606$, $y =
15.9371\,\mathrm{eV}/E$.  In Berkowitz's evaluation, a 4-term polynomial is used, i.e. $a_6 =
a_7 = 0$.  Here, we use a 6-term polynomial for the first range and a 4-term polynomial
for all the rest.

\tablecoeff

For the first range, 15.9371 to 29.3295\,eV, the data of Samson \& Stolte
2002~\cite{samson2002} has become available since Berkowitz's review.  
See \fig{osd3}.  This
motivates refitting, and we have found that a substantially better fit results
from adding two additional terms to Berkowitz's polynomial.  The coefficients
are shown in table~\ref{tab:coeff}.  The region of window resonances from 26.6
to 29.2\,eV is excluded from the fit; these are treated as separate discrete
transitions below.  Samson 1966 and Samson 2002 are used in the fit, but the
Samson 1966 data from 29 to 30\,eV is excluded from the fit for both this
energy range and the following range, as it is in substantial conflict with
other measurements.  The data of Carlson et al 1973~\cite{carlson1973} is not
directly used, but is displayed to illustrate the consistency and to guide the
choice of adopted uncertainty.  Samson 1966 does not state an uncertainty
except to suppose that the overall error on the sum of all continuous
oscillator strengths is about 5\%.  Samson 2002 says that uncertainties range
from 1 to 3\% but unfortunately does not specify which energies have the lower
or higher uncertainties.  Carlson states no uncertainties.  The older two
measurements are largely within 3\% of Samson 2002 and within 3\% of the fit,
with the largest difference between the fit and Samson 1966 being 6\% around
26\,eV and the largest difference for Carlson being 4\% around 19\,eV.  We
adopt 3\% as the uncertainty for this range.  Here and in the following, the
number given indicates the uncertainty on the overall normalization of the
range.

The window resonances evident in \fig{osd3} are 
included as discrete transitions with total oscillator strength
of $-0.055$, following Berkowitz.  Given that two experimental groups,
Madden et al 1969~\cite{madden} and Berrah et al 1996~\cite{Berrah} have results within 10\% of
each other, we assign a 10\% uncertainty.  This is a tiny contributor
to the overall uncertainty on the I-value, and so it is not examined
more closely.

In the following range, 29.3295--48.0\,eV, we performed the fit on the data of
Samson 1966, excluding 29--30\,eV, Carlson 1973, excluding $>$37\,eV where
evidently background dominates, Samson 1991, and Samson 2002.  We note that the
very low point for 43.8\,eV from Samson 1966 is a probable typo --- the
original data table gives 23\,cm$^{-1}$, but 32\,cm$^{-1}$ would be much more
consistent with both the surrounding points and later experimental results.
Since a 6-term polynomial gave very similar results to a 4-term polynomial, the
latter was used; this reasoning holds for each of the following energy ranges
as well. Nearly the same result is obtained if Samson 2002 is fit
alone.  Given the less-good agreement between experiments in this region, a 4\%
uncertainty is adopted.
We note the rather large fractional difference between our evaluation
and that of Berkowitz towards the end of this range.  Our evaluation is
clearly a better fit to all of the data, but in any case the 
\emph{absolute} difference is very small since there is little
oscillator strength between 45 and 48\,eV.

Since both the 15.9371--29.3295\,eV and
29.3295--48.0\,eV ranges are dominated by the work of Samson, their
uncertainties are treated as fully correlated.  The 15.9371--48\,eV range is
the dominant contributor of uncertainty to the I-value, giving $\pm 3.6$\,eV.

\figurefour

For 48.0\,eV to 79.3\,eV, we fit a 4-term polynomial to the data
of Watson 1972~\cite{Watson}, Samson 2002 and  Suzuki \& Saito 2005~\cite{suzuki2005}.
See \fig{osd4}. Watson states 3\% uncertainty.  Suzuki says that uncertainties are between
0.05\% and 2\%, and within 1\% for most energies, but gives no information
as to which energies have which uncertainties.  Accordingly, we assign
2\% uncertainties in the fit.   The three experiments have good
agreement, with only one point of Watson lying more than 2\% from the
fit.  With this robust confirmation from different groups, we adopt a
2\% uncertainty for this range.

From 79.3 to 243\,eV, the data of the same three groups is fit, plus that of
Henke et al 1967~\cite{henke1967}.   This latter data is treated as having
5\% uncertainties, despite the very small errors reported by the authors in this
unpublished report.
There is somewhat more disagreement between the various results in this range,
and so we adopt a 3\% uncertainty.  As nearly all the information comes
from the same groups as the previous energy range, the two are treated
as fully correlated.

\figurefive

From 243 to 250\,eV, the L-edge, we directly use the data of Suzuki 2005 
shown in \fig{osd5}.  From
250 to 336\,eV, we directly average Suzuki 2005 with Chan 1992, without any
functional form, scaled by their stated uncertainties.  Although Chan's data
covers the entire range from the ionization potential to 500\,eV, we do not
display it for the L-edge because with a resolution of 1\,eV, it lacks the
ability to resolve the structure of the edge.  (We have not displayed
it for lower energies because it would add little information.)

From 336 to 500\,eV, we display Chan's data, but do not use it.  Above 336\,eV,
it rises steadily away from the work of other groups, and, as Berkowitz states,
its use in this region would give too much contribution to the sum rule.   We
fit the data of Suzuki to a polynomial, as above.  From 243 to 500\,eV, we assign
an uncertainty of 5\%, given that the evaluation rests heavily on a single
group's data, Suzuki's, and the main second source, Chan, does not agree very well,
even in the 250--336\,eV range.  Two points from Henke 1967 do agree well with Suzuki,
but this is quite sparse.  We also display on the plot the Henke et al
1993~\cite{henke1993} evaluation for comparison only.

\figuresix

From 500 to 929\,eV, we again fit a polynomial to Suzuki's data.  See
\fig{osd6}. In this range
there are four points by Henke 1967 which agree very well, and we choose an uncertainty
of 3\%, fully correlated with the 243--500\,eV range.  For this range and the following,
The evaluation of Chantler et al 1995~\cite{Chantler} is displayed along with Henke's
evaluation, for comparison.

From 929 to 3202\,eV, just before the K-edge, we fit the data of Wuilleumier
1965~\cite{Wuilleumier}, Henke 1967, Suzuki 2005, and Zheng
et al 2006~\cite{zheng2006} to a polynomial.  Given the level of agreement between
these several groups, we assign a 3\% uncertainty, uncorrelated with the previous
regions.

Between 3202 and 3206\,eV, we treat the oscillator strength as consisting of
the discrete 1s$\rightarrow$4pm resonance, measured by Deslattes et al
1983~\cite{Deslattes}, with a strength of 0.0022 and a 10\% uncertainty.  This
is for completeness only, as it has a negligible impact on the I-value or its
uncertainty.

From 3206\,eV to 10\,keV, we fit the data of Wuilleumier 1965, Zheng 2005,
Millar \& Greening 1974~\cite{millar1974} and McCrary et al
1970~\cite{McCrary}.  (Here we depart from direct use of Berkowitz's energy
divisions by using a single fit for this range, which Berkowitz divides into
two sections.) For the higher energy data of
Millar and McCrary, we use the evaluations of Chantler to subtract the
non-photoabsorption portion of the cross section. Again an uncertainty of 3\% is
assigned based on the agreement between groups.  Given the overlap of the experimental groups between the 3206\,eV to 10\,keV range and the 929
to 3202\,eV range, we treat these two ranges as fully correlated.  
This choice is not crucial; a less
conservative choice to consider them uncorrelated would lower the
uncertainty on the I-value by only 0.2\,eV. 

\figureseven

From 10 to 100\,keV, we follow Berkowitz and directly use Chantler's evaluation.
We assign a 3\% uncertainty based on agreement with the data of Millar and McCrary,
being somewhat conservative because of the subtraction procedure necessary to
isolate the photoabsorption component of attenuation at high energies. 
See \fig{osd7}.

Above 100\,keV, we are unaware of any photoabsorption data, and we follow
Berkowitz, using the formula of Bethe \& Salpeter~\cite{Bethe_Salpeter},
directly evaluating it up to 1\,GeV and then integrating the asymptotic form
from there to infinity.  Because the range from 100\,keV to infinity gives
almost no contribution to the I-value, changing it by only 0.1\,eV if it is
neglected entirely, no uncertainty is assigned.

As stated above, the result for the I-value from combining all of these energy
ranges is \garanswerosdonly.  There are several energy ranges each contributing significantly
to the uncertainty on the I-value.  From the ionization potential to
48\,eV has the biggest contribution because of the large amount of
oscillator strength present.
The range from the L-edge to 929\,eV contributes
near the second most because of its fairly large oscillator strength
and larger fractional uncertainties.  Similarly, the range just above the
K-edge is a significant contributor to the uncertainty.  Discrete
transitions below the IP are next most important given their 
substantial contribution to the total oscillator strength and 10\%
uncertainties.  In contrast, the range 48--243\,eV has a very small
contribution because of the fairly small oscillator strengths coupled
with low uncertainties, and above 10\,keV, the contribution to the
uncertainty is small because the oscillator strength is small.

As a byproduct of this calculation, we can estimate the quantities
$I(-1)$ and $I(1)$, where $I(p)$ is defined as:

\[\log I(p) = \left.\left(\sum_n f_n E_n^{p} \log(E_n) +
\int_\mathrm{IP}^\infty \frac{df}{dE} E^{p} \log(E)
dE\right)\middle/S(p)\right.,\] and \[S(p) = \sum_n f_n E_n^{p} +
\int_\mathrm{IP}^\infty E^{p} \frac{df}{dE} dE,\] such that 
the I-value that is the focus of this note is $I(0)$.

We find $I(-1) = 26.51\pm 0.28$.  For $I(-1)$, the lower energies
are more important.  The uncertainty comes almost entirely from discrete
transitions below the IP and from the region just above the L-edge, in
equal proportions.

The result for $I(1)$ is $(3600 \pm 80)\,\mathrm{eV}^2$.  For $I(1)$, higher
energies are more important.  Nearly all of the uncertainty comes from the
region above the L-edge ($\pm 60\,\mathrm{eV}^2$) and from the 10-100\,keV range where we use
the evaluation of Chantler ($\pm 40\,\mathrm{eV}^2$).  This assumes an 
uncertainty below $\sim$10\% above 100\,keV where we use the formula of Bethe and
Salpeter.  Unlike for $I(0)$, the region above 100\,keV is not negligible for $I(1)$;
neglecting it reduces the result by $200\,\mathrm{eV}^2$.
For this reason, we do not venture an estimate for $I(2)$ since it is
even more sensitive to high energies and we are unaware of any 
photoabsorption data above 100\,keV where the majority of the uncertainty
is most likely to lie.

\subsubsection{Comparison with other results}

In 2010, Kumar and Thakkar gave another estimate of argon's I-value, 186.3\,eV
with an estimated $\pm 2$\% ($\pm 3.7\,$eV) uncertainty~\cite{Kumar_Thakkar}.
There are three major differences between their evaluation and ours.  First,
their method uses the sum rule and molar refractivity data as constraints.  We
choose to use oscillator strengths alone, without constraints from molar
refractivity data.    

Second, for any given energy interval, they choose a single data set or
evaluated set of oscillator strengths for their fit.  This has the effect of
discarding modern oscillator strength data in many energy ranges. For instance
from 319.9\,eV to 100\,keV, the 1973 evaluation of Veigele~\cite{veigele1973}
is used even though data from Millar 1974, Suzuki
2005 and Zheng 2006 all exist in this range.
In contrast, we include all modern data, and in each energy range use a fit to all
data judged as reliable.

Third, we display the underlying uncertainties assigned to each part of the
oscillator strength distribution and the effect that each has towards the
final uncertainty on the I-value.
It is not clear how to trace
Kumar and Thakkar estimated uncertainty on the \ivalue back to the underlying
data.

Despite all these differences, the present evaluation arrives at a very similar
result for the \ivalue.  For $I(1)$, Kumar and Thakkar find $3620\,\mathrm{eV}^2$
$\pm 3$\% ($\pm 110\,\mathrm{eV}^2$), also very similar to our result.  For $I(-1)$,
they find 26.53 $\pm 1$\% ($\pm 0.27$), essentially identical to our result.

%
%
%
%
%
%
%
%
%
%

\subsection{Periodic trends}\label{sec:calcper}

As an independent method of estimating the I-value, an interpolation can be
done from nearby elements. First, using the I-values for aluminum, silicon and
calcium, a simple interpolation on a plot of Z vs. I/Z can be made.  The result
is 189\,eV.  I-values certainly do not lie on a smooth curve, so this result
cannot be taken too literally.  Nevertheless, it would be surprising, given the
clear periodic trends, if argon's I-value lay much above 200\,eV or much below
170\,eV, even in the abscense of other methods of evaluation.

\figuretwo

A more sophisticated treatment is given in eq.~4.1 of ICRU-37,
\[I_\mathrm{int}(Z) = \frac{I_c(Z)}{Z_2 - Z_1}\left[\frac{I(Z_1)}{I_c(Z_1)}(Z_2-Z)
+ \frac{I(Z_2)}{I_c(Z_2)}(Z-Z_1)\right],\]
where $Z$ is the atomic number of the element whose I-value is to be interpolated,
$Z_1$ and $Z_2$ are the atomic numbers of the next lower and next higher element
with experimentally determined I-values, $I_c$ indicates a calculated I-value and
a bare $I$ indicates an experimental I-value.  ICRU's choice of calculated I-values
for gasses are the set from Chu \& Powers 1972~\cite{chupowers}, and for solids those
from Ziegler 1980~\cite{Ziegler}, displayed in
ICRU-37 Fig~3.2 and reproduced here in \fig{calci}.  Another set of calculated I-values
are those from Dehmer et
al 1975~\cite{Dehmer} and Inokuti et al 1981~\cite{Inokuti}, where the former covers $Z = 1$--18
and the latter 19--38; these are also shown in \fig{calci}.

Although ICRU does not calculate an interpolated value for argon, as it has
experimental data, the result is 194\,eV, given that
$Z_1 = 14$ (silicon), $I(Z_1) = 173$\,eV, $I_c(Z_1) = 123.5$\,eV,
$Z_2 = 20$ (calcium), $I(Z_2) = 191$\,eV, $I_c(Z_2) = 147.9$\,eV,
and $I_c(Z) = 146.3$\,eV.
This procedure
accounts for the phase of the substance.  If argon is treated as a solid, then
$I_c(Z) = 147.9$\,eV instead, and the result is 196\,eV.

The uncertainties on the experimental
data of silicon and calcium (the nearest elements with experimental data on each side
of argon) contribute 3\,eV to the uncertainty of this interpolation.  The choice of theoretical
input is much more important.  If, instead of the ICRU recommendation, Dehmer 
and Inokuti are used (the unused curve of their Fig.~3.2), the result shifts 20\,eV upwards to 214\,eV.
These inputs are assumed to be worse, but still represent a reasonable calculation, and so
we may qualitatively take the theory uncertainty to be somewhat smaller than the difference.
Therefore, this evaluation includes the I-value interpolated from periodic trends as $(194\pm 12)$\,eV.

\subsection{Hartree-Fock wave functions}

Only two calculations for argon listed in ICRU-90 come with uncertainties
stated by the original authors.  One has such large uncertainties as to be
irrelevant.  The other is Bell et al 1972~\cite{bell}, which uses a method
involving Hartree-Fock wave functions. Since it does not share underlying data
with the OSD calculations, it can be included separately in this note's
evaluation.  Bell's result, as given by the ICRU table, is $(174\pm 3.5)$\,eV.
What Bell actually says is 12.8\,Ry, that ``the predictions of different
representations [agreed] to within 2\% in all cases,'' and that a ``full error
analysis [...] is beyond the scope of the present paper.'' Bell goes on to
compare the calculation for helium with a more sophisticated treatment, finding
that the difference is 6\% and ``Errors in $I$ arising from the Hartree-Fock
approximation are probably similar for the other atoms considered here''.
Since both the 2\% and 6\% errors are relevant, an uncertainty of 11\,eV is
more correct, and is used in the evaluation in this note.

\section{\label{sec:gaseval} Evaluation for gaseous argon}

\figureone

The five direct experimental results are combined with the three indirect methods
to produce this note's evaluation (see
\fig{myeval}).  Each underlying result is represented as a PDF which is the sum
of a Gaussian plus a uniform distribution.  The Gaussian's mean and standard
deviation are the quoted central value and uncertainty of the given result.
The uniform distribution has the range 130--240~eV and a normalization
representing a subjective judgment about how likely the result is to be
incorrect (through any means, e.g., unaccounted for systematic error, incorrect
calculation, typographic error, etc.), typically 5--10\%.  This reflects the
tendency of older papers (and sometimes newer papers) to have results that are
incompatible with each other at many times the stated errors.  The Gaussian is
assigned the remainder of the normalization.

The underlying PDFs are multiplied together to produce the evaluated PDF, which
is integrated from the mean to find the evaluated uncertainty.  The present
evaluation for gaseous argon is \garanswer.

Since the method just described
unfortunately must include subjective estimates of the correctness of past
experiments and calculations, a second method was used to check how robust the
result is.  In this method, it is assumed that exactly one of the input
experiments or calculations is incorrect.  The average is taken with Gaussian
distributions only, but with each input experiment or calculation dropped in
turn, producing several results.  A weighted average is then produced from
these results with the weights set by the uncertainty of the experiment or
calculation that was excluded.  The result of this alternate procedure is
$(187\pm 5)$\,eV, which does not differ significantly from the main procedure.
If our own OSD calculation is exempted from this procedure, the result is 
$(187\pm 4)$\,eV, identical to the main result.

Our answer is ultimately very similar to ICRU-90's, with the same central value and 
an uncertainty just 1\,eV larger.
It has, however, been arrived at through a substantially different process.
Our uncertainty is larger than ICRU-90's for several reasons: First,
the ICRU misread the Martin \& Northcliffe's error as 7\,eV rather than
17\,eV.  Second, the present note's OSD evaluation is \garanswerosdonly, which has a larger
uncertainty than ICRU assigned to the several OSD calculations used in their evaluation.
Finally,
ICRU-90 misunderstood the uncertainty of the Bell calculation, understating it
by a factor of three.

This note's evaluation for gaseous argon is dominated by the OSD calculation presented above.
If it is excluded from the average, the result is instead \garanswerwithoutosd.
If only direct experimental evidence from range and stopping power measurements is used,
the value \garanswerdirectonly.  It can be seen that there is good agreement among the
several methods used to estimate the I-value.

\section{\label{sec:liquideval} Evaluation for liquid argon}

\figurephase

Liquid argon is not just a very dense gaseous argon; there is binding energy associated with
the phase change.  Naively, if electrons are bound more strongly, the stopping power should decrease.
Data on this effect is limited, but instructive.  This effect is predicted to be large for
strongly bound systems such as metals, and smaller for molecular substances.  Examples of the
latter:

\begin{itemize}

\item{Hydrogen:} ICRU-37 (Table 5.7) recommends 19.2\,eV for gaseous $\mathrm{H_2}$ and 21.8\,eV for liquid hydrogen (14\% higher).

\item{Water:} The ICRU-90 recommended value for liquid water is $(78\pm 2)$\,eV.  ICRU-37 gives
$(71.6\pm 2)$\,eV for water vapor and $(75\pm 3)$\,eV for liquid water.   The addendum for ICRU-73~\cite{icru73add}
gives 69.1\,eV for water vapor and 78\,eV for liquid water.  ICRU-90 cites this without recommending a
value for water vapor.  The phase effect on the I-value ranges from 5\% to 13\% depending on the numbers chosen. 
For the purposes of this evaluation, the phase effect is considered to be $(9\pm 4)$\%.

\item{Nitrogen:} ICRU-37 recommends $(82\pm 2)$\,eV for gaseous $\mathrm{N}_2$ and $(90.5\pm 2.6)$\,eV for liquid nitrogen. After converting their 90\% C.L. errors to standard ones, this gives an increase of $(10\pm 3)$\%.

\item{Oxygen:} ICRU-37 recommends $(95\pm 2)$\,eV for gaseous $\mathrm{O}_2$ and $(104.3\pm 2.6)$\,eV for liquid oxygen ($(10\pm 3)$\%).

\item{n-propane, n-pentane, n-hexane, n-heptane: } ICRU-37 recommends I-values differing by 10\% or 11\% between the
liquid and gas phases.  All of the gaseous measurements were performed by the same group, as were all of the
liquid measurements, so the uncertainties are considered fully correlated. Averaging them gives $(11\pm 5)$\%.

\item{Bromine: } ICRU-37 recommends 343\,eV for gaseous bromine and 357\,eV for condensed bromine, although
both are interpolations from adjacent elements.  Taking into account the uncertainties of this procedure as
was done for argon in \sect{calcper}, this is considered to be a change of $(4\pm 6)$\%.

\item{Iodine: } ICRU-37 recommends 474\,eV for gaseous iodine and 491\,eV for condensed iodine.
Using the same procedure as for bromine, an evaluation of $(4\pm 6)$\% is obtained. 

\end{itemize}

None of these are noble gasses like argon, but several have similar boiling points and are likewise non-polar, which implies that
the relevant binding energies are similar.  The effect decreases with atomic number, as expected, since a smaller fraction of
the electrons participate in chemical binding.  By fitting a smooth curve to the changes as a function of $Z$ (see \fig{interp}), we would expect around a 7\% increase in the I-value of argon.  An uncertainty of 3\% is estimated, based on the mean of the
uncertainties for water, nitrogen and oxygen, the closest three experimental data points.

There is a limited amount of experimental evidence for argon itself. As
summarized in ICRU-49~\cite{icru49}, two groups have studied the difference of
stopping power between solid and gaseous argon, but unfortunately only with
alpha particles below 3\,MeV.  Chu et al 1978~\cite{chu1978} found a 5--10\%
decrease in stopping power for solid argon below 1.0\,MeV, and none for
1--2\,MeV.  Besenbacher et al 1981~\cite{Besenbacher1981} found no phase effect
to within their 3\% uncertainty for 0.5--3.0\,MeV alphas, implying less than a
5\% change in the mean excitation energy.  Moreover, Besenbacher 1981 points
out that their measurements of solid argon are compatible with Chu 1978; the
apparent difference comes from the use of different values for the gaseous
stopping power.  The situation remains a little confused, but the result from
Besenbacher alone still allows for a change of the I-value of up to 9\,eV.
This is surprisingly small compared to what's expected from the results from
the non-polar molecules listed above.  It may be that results from slow alpha
particles do not reliably translate to results for fast protons.

Solid argon likely has a very similar stopping power to liquid argon, as both
are condensed phases.  Unfortunately, to the author's knowledge, there are no
substances for which the I-value has been measured in both solid and liquid
phases. This lack of experimental evidence motivates conservatism in the error
assignments.

Since data from other molecules suggests an increase in the I-value of $(7\pm 3)$\%,
while data from solid argon suggests
$(0\pm 5)\%$, it seems reasonable to take the weighted average of these and arrive at an estimated
I-value increase of $(5\pm 3)$\% for argon in a condensed phase.  This gives
\laranswer as the present evaluation for the mean excitation energy of
liquid argon.

\section{\label{sec:implications} Implications}

Geant4~\cite{Geant}, by default, uses the (older) ICRU-37 central value for gaseous argon,
188\,eV, regardless of phase.  In a liquid argon neutrino detector, the two
most important quantities that the I-value feed into are muon $dE/dx$ at minimum
ionization and muon range.  The former is used for calorimetric calibration and
the latter to measure muon energy.

At muon minimum ionizing, around 270\,MeV, changing the I-value from 188\,eV to
197\,eV decreases $dE/dx$ by 0.3\%.  This shift increases towards lower energy
and is 0.5\% at 100\,MeV.  The effect on calibration is dependent on the details of the
procedure, but is perhaps 0.4\%.  Muon range at 1\,GeV is increased by 0.3\%.  As with $dE/dx$, this
shift increases as muon energy decreases, and is 0.6\% at 100\,MeV.

As examples of two other quantities that affect simulation and reconstruction,
the effect on proton $dE/dx$ and range is also calculated.  Proton $dE/dx$ is
decreased by 0.7\% at 50\,MeV and 0.6\% at 400\,MeV. Across these energies,
range increases by between 0.6\% and 0.7\% with the largest change around
100\,MeV.

Since the change in the I-value is 9\,eV and the recommended uncertainty is
7\,eV, the uncertainties on all these ranges and stopping powers are nearly as
large as the shifts quoted above.  Notably, the effect of a change in I-value on calorimetric energy
calibration with muon $dE/dx$ and the muon energy reconstruction with range have
the same sign: the whole neutrino event energy is overestimated by using
gaseous argon's I-value.  Nor is there any cancellation of this kind of uncertainty
from use of a liquid argon near detector.

Energy reconstruction has many uncertainties besides those from the I-value, but the
author's estimate is that the I-value dominates muon energy estimation uncertainty below
1\,GeV.  Most of the information about $\nu_\mu$ energy comes from the muon, and 
$\Delta m^2_{32}$ is directly proportional to the measured energies of oscillation
minima and maxima.  All other oscillation parameters rest, if not as directly, on energy
reconstruction as well.   The I-value is, therefore, a critical parameter for DUNE and other
liquid argon detectors.

To conclude, this note recommends a new value and uncertainty for the mean
excitation energies of gaseous and liquid argon, \garanswer and \laranswer,
respectively.  The central value for liquid argon is significantly higher than
that most recently recommended by the ICRU for gaseous argon, and the
uncertainty is substantially larger.  While this recommendation is believed to
be a useful improvement, it is notable that it rests strongly on an indirect
calculation based on oscillator strength distributions.  Direct experimental
evidence is also used, but none of the inputs are clean.  Three of the five
experiments lack any original statement of uncertainty, while the remaining two
give numerical uncertainties, but without confidence levels.  Of these latter
two, only one directly states an I-value.  With so much freedom of
interpretation, another evaluation could easily arrive at a different result.
This fact should further motivate experimental work to improve the uncertainty.

\section{Acknowledgments}

This work was supported by the U.S. Department of Energy.  Thanks also to
Andrew Furmanski, Tom Junk, and Abigail Waldron for helpful discussions.

\bibliographystyle{JHEP}
\bibliography{larIrecpaper.bib}

\end{document}